\pdfoutput=1
\documentclass[reprint, amsmath, amssymb, aps, unsortedaddress,superscriptaddress]{revtex4-2}
\usepackage{graphicx}
\usepackage{float}
\usepackage{dcolumn}
\usepackage{color}
\usepackage{latexsym,bm}
\usepackage[normalem]{ulem}
\usepackage{multirow}
\usepackage{appendix}
\usepackage{amsmath}
\usepackage{amsfonts}
\usepackage{booktabs}
\usepackage{array}
\usepackage{soul}
\usepackage{CJKutf8}
\usepackage{bm}

\newcommand{\QL}[1]{\textcolor{black}{{#1}}}
\newcommand{\MC}[1]{\textcolor{black}{{#1}}}

\begin{document}
\hbadness=10000
\hyphenpenalty=5000
\tolerance=1000
\begin{CJK}{UTF8}{gkai} 
 
\title{Thermal and Electrical Conductivities of Aluminum Up to 1000 eV: A First-Principles Prediction}

\author{Qianrui Liu (刘千锐)}
\affiliation{HEDPS, CAPT, College of Engineering and School of Physics, Peking University, Beijing 100871, P. R. China}

\author{Xiantu He (贺贤土)}
\affiliation{HEDPS, CAPT, College of Engineering and School of Physics, Peking University, Beijing 100871, P. R. China}
\affiliation{Institute of Applied Physics and Computational Mathematics, Beijing 100088, China}

\author{Mohan Chen (陈默涵)}
\email{mohanchen@pku.edu.cn (Corresponding author)}
\affiliation{HEDPS, CAPT, College of Engineering and School of Physics, Peking University, Beijing 100871, P. R. China}

\date{\today}

\begin{abstract}
{
Accurate prediction of the thermal and electrical conductivities of materials under extremely high temperatures is essential in high-energy-density physics. These properties govern processes such as stellar core dynamics, planetary magnetic field generation, and laser-driven plasma evolution.
However, first-principles methods like Kohn-Sham (KS) density functional theory (DFT) face challenges in predicting these properties due to prohibitively high computational costs.
We propose a scheme that integrates the Kubo formalism with a mixed stochastic-deterministic DFT (mDFT) method, which substantially enhances efficiency in computing thermal and electrical conductivities of dense plasmas under extremely high temperatures. As a showcase, this approach enables {\it ab initio} calculations of the thermal and electrical conductivities of Aluminum (Al) up to 1000 eV.
Compared to traditional transport models, our first-principles results reveal significant deviations in the  thermal and electrical conductivities of Al within the warm dense matter regime, underscoring the importance of accounting for quantum effects when investigating these transport properties of warm dense matter.
}
\end{abstract}

\maketitle
\end{CJK}

{\it Introduction}---Accurate prediction of thermal and electrical conductivities in dense plasmas is crucial for advancing research in astrophysics and inertial confinement fusion (ICF). 
Thermal conductivity is critical for studying the structures and evolution of planetary cores~\cite{68AJ-Hubbard,24SA-Hsieh, 25NC-Wang}.
Additionally, it plays a significant role in the development of ablator materials and internal fuel capsules~\cite{98PP-Marinak,14E-Hu,20MRE-Kang}, as well as in the interactions between laser and targets~\cite{03B-Ivanov}, which are important for designing ICF experiments.
Electrical conductivity, on the other hand, plays a key role in understanding the metallization of high-pressure hydrogen~\cite{96L-Weir,99B-Nellis,09L-Lorenzen} and helium~\cite{81B-Young,07L-Kietzmann,10L-Celliers,09L-Lorenzen}, which is essential for modeling the interiors of gas giants and brown dwarfs.

Measuring thermal conductivity experimentally remains challenging.~\cite{17SR-McKelvey, 23CP-Jiang, 25NC-Allen} For instance, determining the thermal conductivity of dense plasmas is severely constrained by their extreme conditions and short lifetimes. 
While the Wiedemann-Franz law enables direct estimation of thermal conductivity from electrical conductivity in certain low-temperature experiments,~\cite{98RSI-Rhim, 07IJT-Brandt} this relationship breaks down for dense plasmas due to the temperature dependence of the Lorentz number.~\cite{17SR-McKelvey}

Furthermore, despite the existence of various simulation models designed to describe thermal and electrical conductivities of dense plasmas, such as Spitzer~\cite{53PR-Spitzer, 06-Spitzer}, Lee-More~\cite{84PF-Lee}, and Ichimaru~\cite{85A-Ichimaru}, the accuracy of these models is still limited. 
The first-principles approach based on Kohn-Sham Density Functional Theory (KSDFT)~\cite{64PR-Hohenberg,65PR-Mermin,65PR-Kohn} and the Kubo-Greenwood formula~\cite{57JPSJ-Kubo,58PPS-Greenwood} has been widely used for predicting the conductivities of materials~\cite{02E-Desjarlais, 05B-Recoules,12B-Vlcek,18PP-Witte,21MRE-Liu}.
However, KSDFT faces significant challenges at high temperatures due to the rapid, cubic ($O(T^3)$) increase in computational cost with temperature~\cite{18B-Cytter}.
In this regard, quantum Monte Carlo (QMC) was proposed as an alternative first-principles method for calculating the electrical conductivity around a decade ago~\cite{09L-Lin}, 
Moreover, although methods based on the average-atom (AA) model have been proposed to cover a broader temperature range for dense plasmas~\cite{19E-Faussurier, 20E-Shaffer, 20E-Wetta,21E-Rightley}, their accuracy requires rigorous validation against first-principles results.

The mixed stochastic-deterministic density functional theory (mDFT)~\cite{20L-White} is a promising alternative to conventional KSDFT for high-temperature systems.
It combines the precision of KSDFT with the high-temperature efficiency of stochastic density functional theory (sDFT)~\cite{13L-Baer,18B-Cytter}, achieved by using stochastic wave functionals rather than explicitly diagonalizing the Hamiltonian.
Compared with sDFT, mDFT employs deterministic Kohn-Sham orbitals to describe low-energy, tightly bound states that are crucial for accuracy yet inexpensive to treat, thereby substantially reducing stochastic errors.
While sDFT-based methods for electrical conductivity have been developed~\cite{19B-Cytter}, their application to dense plasmas has remained limited, primarily due to non-negligible stochastic noise. Moreover, to the best of our knowledge, no efficient first-principles methods for calculating thermal conductivity have been reported to date, despite it being more challenging to measure experimentally.

We present a unified first-principles methodology within the mDFT framework for computing both electrical and thermal conductivities from Onsager coefficients across a wide temperature range, including the hot dense plasma regime.
Our method yields results in close agreement with those obtained using the Kubo-Greenwood approach of KSDFT, while offering greater efficiency for high-temperature systems.
More details can be found in the supplementary material~\cite{sm}. 
We apply this approach to calculate the electrical and thermal conductivities of aluminum \MC{(Al)} over a broad temperature range, extending first-principles calculations from the few-tens-of-eV limit of conventional KSDFT to 1000 eV, where electrons are nearly fully ionized. 

\paragraph*{Methods.} 
Both sDFT and mDFT methods do not directly yield eigenvalues, and instead, they compute physical properties based on the trace of the corresponding quantum mechanical operators. In this context, conductivities can be derived from the current response function, as described by the Kubo formula~\cite{57JPSJ-Kubo}, where the response function is expressed in terms of the trace of the current operator
\begin{equation}
    \label{eq:current_response}
        C_{nm}(t) = -\frac{i}{\hbar}\theta(t)\mathrm{Tr}\left\{\hat{f}_H\left[\hat{\mathbf{j}}_n, \hat{\mathbf{j}}_m(t)\right]\right\},
\end{equation}
where \(\bigl[\, , \,\bigr]\) denotes the commutator, $\theta(t)$ is the Heaviside step function, and $\hat{f}_H$ is the Fermi-Dirac function of the Hamiltonian operator $\hat{H}$, which represents the density matrix $\hat{\rho}$ in the single-particle system. 
The operator $\hat{\mathbf{j}}_1$  and $\hat{\mathbf{j}}_2$ represent the electrical and heat current, respectively.

To calculate conductivities, we employ mixed KS-stochastic orbitals to evaluate the trace in the current response function shown in Eq.~(\ref{eq:current_response}), rather than relying solely on stochastic orbitals. This hybrid strategy reduces stochastic errors and improves efficiency, particularly when treating tightly bound core states. In addition, representing low-energy core electrons with KS orbitals raises the minimum energy in the stochastic subspace, thereby narrowing the spectral width and substantially decreasing the order required in the Chebyshev expansion used for time propagation, which further enhances computational efficiency. Since existing stochastic DFT formulations~\cite{19B-Cytter} are not directly applicable to mixed KS-stochastic orbitals, we derived a new trace expression (see supplementary material~\cite{sm} for details), given by 
\begin{equation}
\begin{aligned}
    T_{nm}(t) = &\sum_{a,b} \langle\varphi_a^+(\tfrac{t}{2})|\hat{\mathbf{j}}_n|\varphi_b^-(\tfrac{t}{2})\rangle \\
    &\times \langle\varphi_b^-(-\tfrac{t}{2})|\hat{\mathbf{j}}_m|\varphi_a^+(-\tfrac{t}{2})\rangle,
\end{aligned}
\end{equation}
where $T_{nm}$ is the trace part in Eq.~(\ref{eq:current_response}) and
the time-evolved states are defined by
\begin{equation}
\begin{aligned}
    |\varphi^+(\tfrac{t}{2})\rangle &\equiv e^{i\hat{h}t/2}\hat{f}_H^{1/2}|\varphi\rangle, \\
    |\varphi^-(\tfrac{t}{2})\rangle &\equiv e^{i\hat{h}t/2}(1-\hat{f}_H)^{1/2}|\varphi\rangle,
\end{aligned}
\end{equation}
with $\hat{h}=\hat{H}/\hbar$ and $|\varphi\rangle$ denoting KS or stochastic orbitals. For KS orbitals, the above expressions reduce to simple substitutions of their eigen energies, avoiding any additional computation.

We performed molecular dynamics (MD) simulations of aluminum at a density of 2.7 g/cm$^3$ and a temperature range of 10-1000 eV with the ABACUS package~\cite{10JPCM-Mohan,16CMS-Li,22B-Liu}. Simulation cells contained 32 atoms for temperatures up to 30 eV, 16 atoms for temperatures up to 200 eV, and 4 atoms for higher temperatures. Each simulation was run for 2000 steps with a time step of $\frac{r_s}{60\bar{v}}$, where $r_s$ is the Wigner-Seitz radius and $\bar{v}$ is the average velocity. The temperature was controlled by the Nosé-Hoover thermostat~\cite{84JCP-Nose,85A-Hoover}. Since the temperature-dependent exchange-correlation (XC) functional has only a minor impact on the results~\cite{24B-Liu}, we adopted the Perdew-Burke-Ernzerhof (PBE) exchange-correlation functional~\cite{96L-Perdew}, with a $2\times2\times2$ Brillouin zone sampled using the Monkhorst-Pack method~\cite{76B-Monkhorst}. 
We then selected 5 configurations to calculate the current response functions using the approach outlined in Eq.~(\ref{eq:current_response}).
To ensure accuracy, we used pseudopotentials with 13 valence electrons for calculating conductivities at $T>10$ eV, while at 10 eV results with 11 and 13 valence electrons agree well.

\begin{figure}
	\centering
	\includegraphics[width=8.6cm]{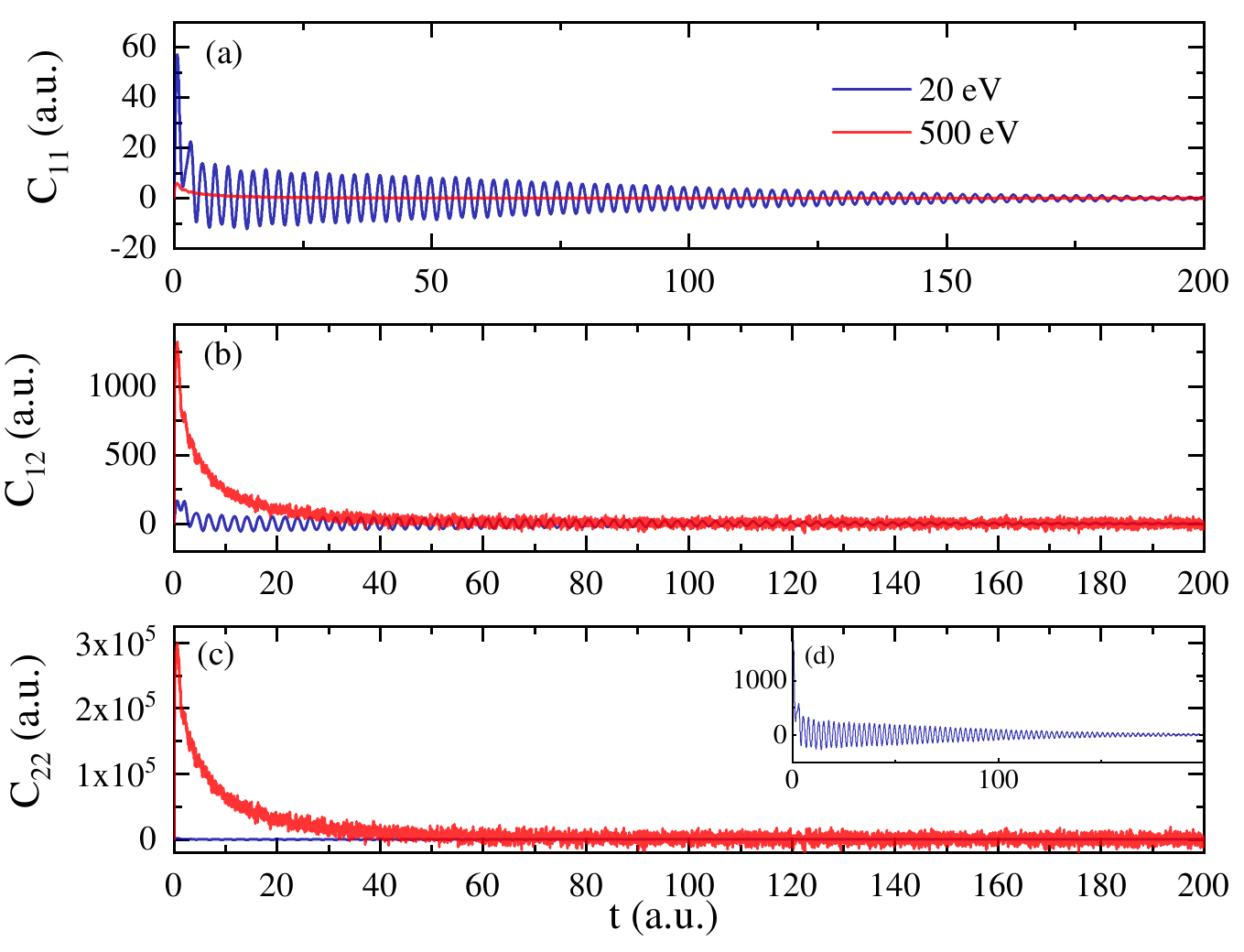}\\
	\caption{(Color online) 
    Current response functions $C_{mn}$ of Aluminum, at $T=20~\mathrm{eV}$ (blue) and  $T=500~\mathrm{eV}$ (red), including (a) $C_{11}$, (b) $C_{12}$, and (c-d) $C_{22}$. Panel (d) shows the same data as in (c) but with the vertical scale enlarged by 100× for clarity.
    }
	\label{fig:jj}
\end{figure}

\paragraph*{Results and Discussion.} 
Fig.~\ref{fig:jj} shows the current response functions $C_{mn}$ of aluminum at $T =20$ and 500 eV. $C_{mn}$ curves exhibit oscillatory structures, with notable differences attributed to different energy levels. 
At low temperature (20 eV), electron transitions occur within a narrow energy range, resulting in narrow-spectral oscillations.
At high temperature (500 eV), the introduction of higher-energy states leads to broadband excitation, causing the oscillations to take on a noise-like character. 
Furthermore, for the response of electrical current $C_{11}$, the calculation at 20 eV produces higher magnitudes. However, for that of heat current $C_{22}$, the inclusion of higher energy levels at 500 eV results in higher magnitudes. 
Additionally, the response function decays more rapidly towards zero at lower temperatures, implying a smaller relaxation time. 
These behaviors highlights the different effects of energy levels and temperatures on the characteristics of the response function.

\begin{figure}
	\centering
	\includegraphics[width=8.6cm]{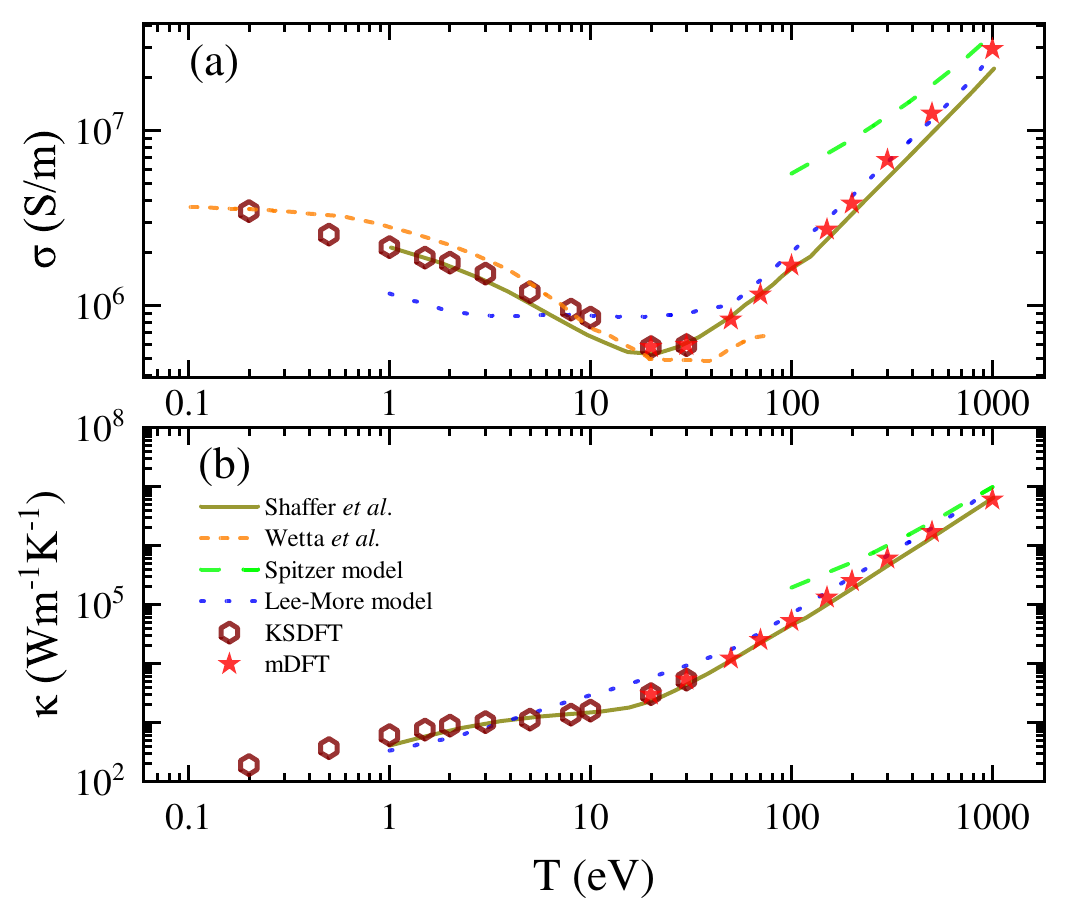}\\
	\caption{(Color online) (a) Electrical conductivity $\sigma$ and (b) thermal conductivity $\kappa$ of Al from 0.2 to 1000 eV. Shown for comparison are traditional plasma transport models, including the Spitzer model~\cite{53PR-Spitzer} (green line) and the Lee-More model~\cite{84PF-Lee} (blue line). Results based on average-atom models are also plotted: Shaffer {\it et al.}~\cite{20E-Shaffer} (yellow-gray line) and Wetta {\it et al.}~\cite{20E-Wetta} (orange line). First-principles calculations are given by KSDFT (dark red open hexagon) and mDFT (red stars). For $T < 10$ eV, KSDFT results are taken from our previous work~\cite{24B-Liu}.}

	\label{fig:cond}
\end{figure}

Fig.~\ref{fig:cond} illustrates the {\it ab initio} electrical and thermal conductivities of Al up to 1000 eV.
Results from mDFT (red solid stars) and KSDFT (dark red hollow circles) are in close agreement, establishing the accuracy of mDFT for transport calculations. The electrical conductivity decreases with temperature before rising again, with a minimum near 20 eV, which is associated with the onset of $L$-shell ionization and the thermal excitation of electrons to higher-energy states (see Fig.~\ref{fig:cv} and discussion therein).
In contrast, the thermal conductivity increases monotonically, with a change in slope around 20 eV.

\MC{A comparison with existing models shows a marked discrepancy in their agreement.} The effective charges used in both Spitzer and Lee-More models in this comparison are obtained from KSDFT and mDFT calculations. Although the Lee-More model~\cite{84PF-Lee} captures the characteristic minimum in electrical conductivity, its accuracy at lower temperatures differs markedly from our first-principles results, and the position of this minimum also varies. When temperatures are above 200 eV, its results agree well with mDFT results.
Conversely, since the Spitzer model~\cite{53PR-Spitzer} assumes full ionization, its results deviate from ours at lower temperatures but begin to converge as the temperature increases.
Average-atom (AA) models often incorporate adjustable parameters and model-dependent choices, leading to varied outcomes. For instance, Shaffer {\it et al.}'s model~\cite{20E-Shaffer}, which integrates the quantum Landau-Fokker-Planck equation and mean-force scattering, generally reproduces the first-principles results well, despite some minor differences in detail. In contrast, Wetta {\it et al.}'s model~\cite{20E-Wetta}, which combines the Ziman formalism, yields results that are notably lower than Shaffer's at high temperatures.
This underscores that different average-atom models, due to their inherent adjustable parameters, can produce inconsistent and sometimes significantly divergent outcomes. Therefore, our parameter-free first-principles calculations are crucial. They provide invaluable, robust reference data, establishing a new standard for validating and refining future modeling efforts in this field.

\begin{figure}
	\centering
	\includegraphics[width=8.6cm]{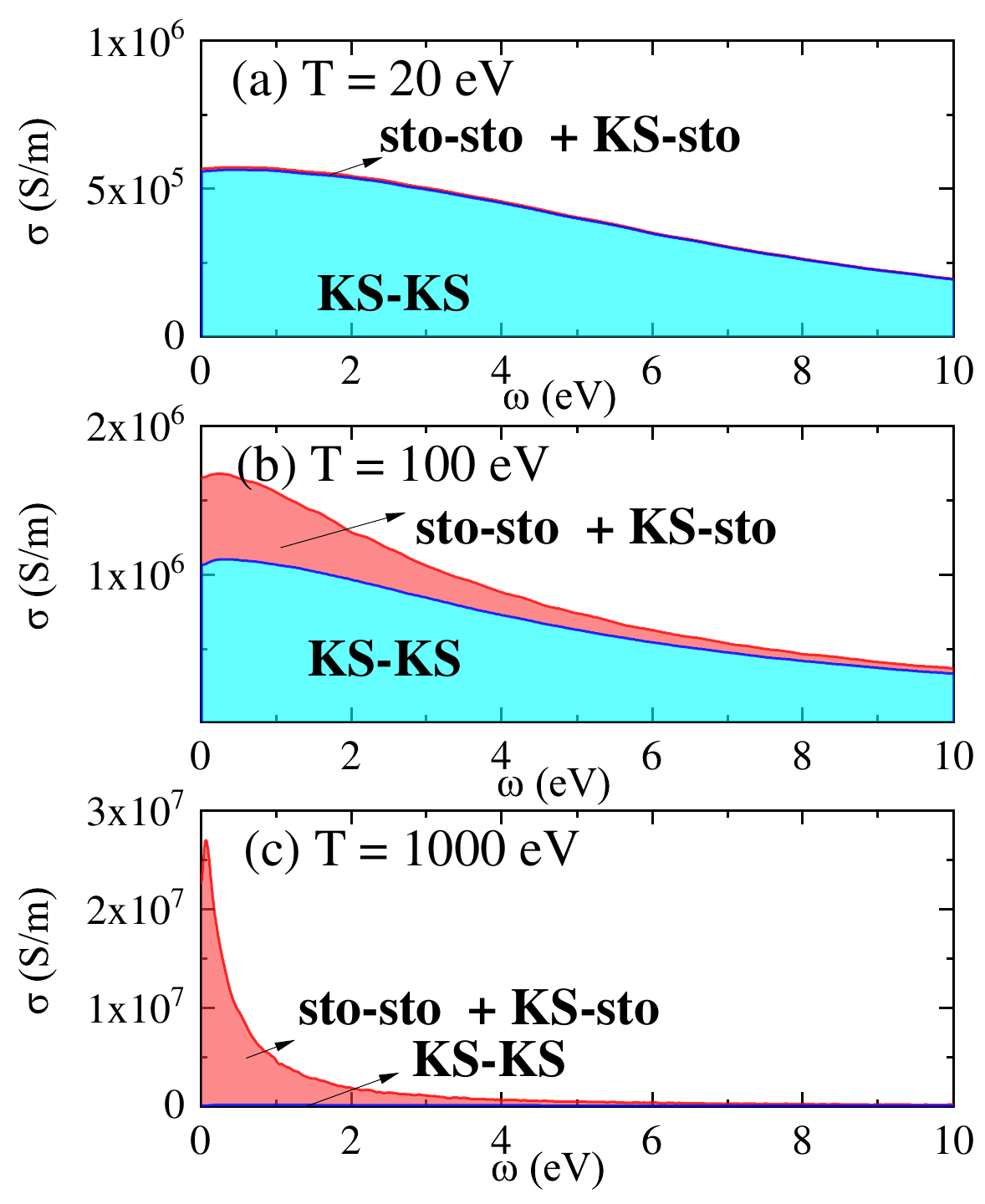}\\
	\caption{(Color online) Contributions to \MC{electrical} conductivities of different orbitals at temperatures of (a) 20 eV, (b) 100 eV, and (c) 1000 eV. Blue regions represent the contributions from pure KS orbitals (KS-KS). Red regions denote the additional contributions from mixed stochastic orbitals, which include both transitions between KS and stochastic orbitals (KS-sto) and transitions among stochastic orbitals themselves (sto-sto).
    }
	\label{fig:contribute}
\end{figure}

Fig.~\ref{fig:contribute} illustrates the ability to calculate high-temperature properties. Stochastic orbitals are effective in capturing the electronic information of high-energy levels. At low temperatures, the contribution from high-energy states is negligible (as shown in Fig. 3(a)), allowing a limited set of KS orbitals to provide converged results. However, as the temperature increases, the contribution from high-energy states becomes more significant. For example, at 100 eV, the contributions from the finite KS orbitals and the high-energy states are comparable. As the temperature rises to 1000 eV, the contribution from the finite KS orbitals becomes almost negligible, making stochastic orbitals crucial.
This highlights the inefficiency of relying solely on KS orbitals at such extreme temperatures. 
Despite having only 96 stochastic orbitals, Fig. 3(c) shows that they capture all the high-energy information. While there are some stochastic errors (which are small, as indicated in the supplementary material~\cite{sm}), the results remain highly accurate.

\begin{figure}
	\centering
	\includegraphics[width=8.6cm]{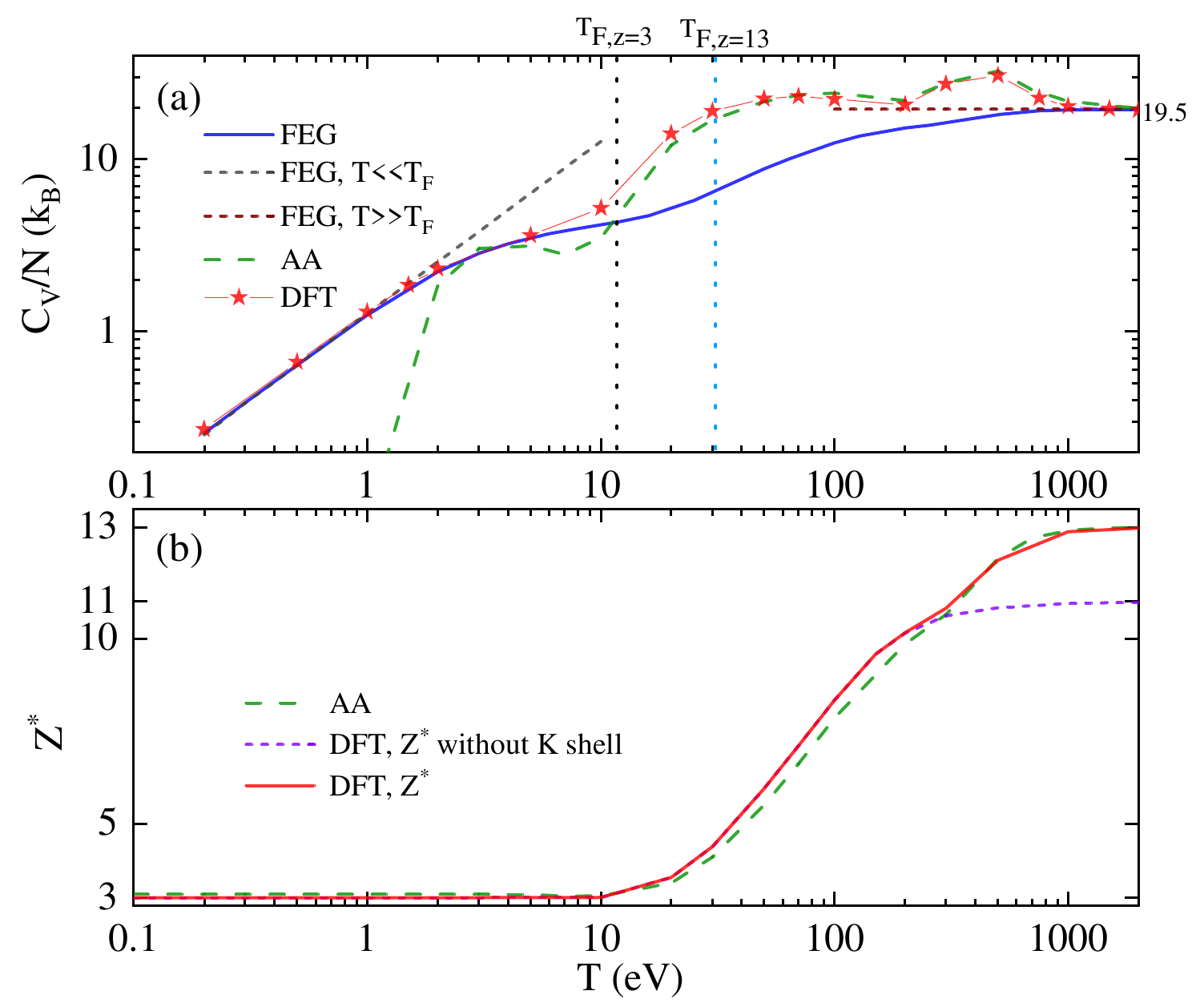}\\
	\caption{(Color online) (a) Electronic isochoric heat capacity and (b) effective charge of Aluminum at different temperatures. ``FEG" denotes results from the free-electron gas model, with $T \ll T_F$ corresponding to the strongly degenerate limit and $T \gg T_F$ to the weakly degenerate limit. ``AA" refers to average-atom model results~\cite{23R-Callow}. ``DFT" results are obtained from KSDFT at $T \leq 10$ eV and from mDFT at higher temperatures.
    }
	\label{fig:cv}
\end{figure}

Using the Drude model, the relationship for thermal conductivity is given by
\begin{equation}
    \kappa = \frac{\tau}{3} C_V \bar{v}^2,
\end{equation}
which leads to the concept of specific heat at constant volume (\(C_V\)). Here $\tau$ is the relaxation time. In the regime where \(T \ll T_F\) (strong degeneracy), the specific heat is proportional to the temperature, and therefore Lorentz number remains constant, which is $\frac{\pi^2}{3}\left(\frac{k_\mathrm{B}}{e}\right)^2$.  However, as the temperature approaches \(T_F\), higher-order temperature terms start to appear in the specific heat, causing it to decrease. This results in a reduction of \(\frac{\kappa}{\sigma T}\), a phenomenon widely observed in previous computational studies.
For temperatures much greater than \(T_F\) (weak degeneracy), the specific heat follows the classical result of the equipartition theorem, \(\frac{3}{2} \cdot 13\), and aluminum, as a dense metallic system, adheres to this result at both high and low temperatures.

Fig.~\ref{fig:cv}(b) shows that the first noticeable rise in the effective charge $Z^*$ occurs near $T \approx 10$~eV, corresponding to the ionization of the $L$-shell electrons. 
This process increases the density of conduction electrons and reduces the core-screening of the ionic potential, which in turn modifies the electron-ion scattering strength (See supplementary material~\cite{sm}). The combined effects lead to the observed minimum and subsequent upturn in the electrical conductivity around this temperature (Fig.~\ref{fig:cond}(a)). 
A second, more abrupt increase in $Z^*$ appears near $T \approx 200$~eV, signaling the onset of K-shell ionization. 
At the ionization onset of both shells, a noticeable inflection appears in the heat capacity $C_V$ (Fig.~\ref{fig:cv}(a)), as a large amount of internal energy is absorbed in liberating tightly bound electrons. 
At $T \approx 1000$~eV, the ionization is essentially complete and the electrons are weakly degenerate, marking a transition beyond the warm dense matter regime.  
For comparison, the AA model~\cite{23R-Callow} qualitatively reproduces the shell-ionization signatures in $Z^*$ and $C_V$ at high temperatures, where electrons are nearly fully ionized. However, it fails to capture the correct shape and location of these transitions, particularly in the low- and intermediate-$T$ regimes where bound-state effects are significant. This limitation stems from its reliance on the Thomas--Fermi description of the electronic structure, which lacks discrete shell resolution.

\paragraph*{Conclusion.}
\MC{We introduced an \QL{efficient approach that combines the Kubo formalism with mixed stochastic-deterministic DFT framework}, enabling efficient calculations of electrical conductivity and thermal conductivity.} Our results demonstrate that this approach is efficient for calculating the conductivity of high-temperature, dense aluminum up to 1000 eV. The conductivity and thermal conductivity values we computed align well with the results from low-temperature KS-DFT calculations. 
This work extends the applicability of first-principles conductivity calculations by two orders of magnitude in temperature, establishing a parameter-free benchmark for extreme warm dense matter and hot dense plasma.

The work is supported by the National Key R$\&$D Program of China under Grant No.2025YFB3003603.
%
The numerical simulations were performed on the High-Performance Computing Platform of CAPT.

\bibliography{skg}

\end{document}